\documentclass[journal]{IEEEtran}
\usepackage{blindtext}
\usepackage{graphicx}
\usepackage{amsmath}
\usepackage{amssymb}
\usepackage{amsbsy}
\usepackage{amsmath}
\usepackage{amsthm} 
\usepackage{cleveref}
\usepackage{caption}
\usepackage{subcaption}
\usepackage{array}
\usepackage{float}
\usepackage{multirow}
\usepackage[linesnumbered,ruled]{algorithm2e}

\usepackage{tabularx}
\usepackage{booktabs}
\usepackage{makecell}

\usepackage{scalerel}
\usepackage{adjustbox}
\def\decibel{\scalerel{$d$\kern-1.4pt}{\clipbox{2pt 0pt 0pt 0pt}{B}}}

\newtheorem{myrmk}{Remark}
\ifCLASSINFOpdf
\else
\fi
\begin{document}
%
\title{Methods of Adaptive Signal Processing on Graphs Using Vertex-Time Autoregressive Models}
%
%
%

\author{Thiernithi~Variddhisai,~\IEEEmembership{Student
                                                Member,~IEEE,}
        and~Danilo~Mandic,~\IEEEmembership{Life~Fellow,~IEEE}
    }

%
%

\markboth{Journal of \LaTeX\ Class Files,~Vol.~6, No.~1, January~2007}%
{Variddhisai \MakeLowercase{\textit{et al.}}: Draft Paper for IEEE Journals}
%



\maketitle
\begin{abstract}
The concept of a random process has been recently extended to graph signals, whereby random graph processes are a class of multivariate stochastic processes whose coefficients are matrices with a \textit{graph-topological} structure. The system identification problem of a random graph process therefore revolves around determining its underlying topology, or mathematically, the graph shift operators (GSOs) i.e. an adjacency matrix or a Laplacian matrix. In the same work that introduced random graph processes, a \textit{batch} optimization method to solve for the GSO was also proposed for the random graph process based on a \textit{causal} vertex-time autoregressive model. To this end, the online version of this optimization problem was proposed via the framework of adaptive filtering. The modified stochastic gradient projection method was employed on the regularized least squares objective to create the filter. The recursion is divided into 3 regularized sub-problems to address issues like multi-convexity, sparsity, commutativity and bias. A discussion on convergence analysis is also included. Finally, experiments are conducted to illustrate the performance of the proposed algorithm, from traditional MSE measure to successful recovery rate regardless correct values, all of which to shed light on the potential, the limit and the possible research attempt of this work.
\end{abstract}

\begin{IEEEkeywords}
Random graph signal, vertex-time stochastic processes, system on graphs, graph shift operator, adaptive graph signal processing, multivariate statistical models, stochastic gradient projection
\end{IEEEkeywords}

%
\IEEEpeerreviewmaketitle

\section{Introduction}
The field of adaptive signal processing has found success in a vast number of applications, from MIMO communication through to real-time machine learning~\cite{Chambers2001,Diniz2012,Tao2013}. Its adoption in less conventional data structures has also been growing with many recent results in quaternions~\cite{Yim2017,Took2010} and tensors~\cite{Yim2019}. The least mean square (LMS) algorithm~\cite{Widrow1959} has been the first fundamental adaptive filtering strategy in all these domains, and the rapidly growing interest in adaptive filtering of multivarite/multiway data types comes as a consequence of the increasing availability of multisensor/multinode data acquisition devices. Typically, the measurement is obtained from a large-scale sensor array, with possibly sparse and arbitrarily distributed sensors which provide streaming data. These challenges require us to move further the traditional methods and problem formulation of adaptive signal processing and to introduce domain-specific solutions.

This paper considers the general scenario of irregular sensor structures represented as a graph, whereby the underlying statistical model follows graph-topological structure. The existing work in signal processing on graphs includes the tracking the time-varying graph signals~\cite{Lorenzo2016,Lorenzo2018} and the use of space transforms and dimensionality reduction to reduce the problem complexity~\cite{Qiu2017,Romero2017}. These results assume an autoregressive model for graph data, and have recently been generalized to autoregressive moving average models~\cite{Isufi2019}.

It is important to note that the common assumption made in much of the existing work is that the graph topology is known \textit{a priori}. However, in many network-related problems, like social media, financial assets, or neuron connectivity, the topology (relationship between nodes, assets or neurons) needs to be learned, not to mention that the topology is also often time-varying. To discover the topology of the GSO which generated the observed graph data, the work in~\cite{Mei2017} assumes a vertex-time autoregressive casual process; however, like all above  mentioned articles, this offers a batch method where all the data are considered at once. To the best of the authors' knowledge, a \textit{truly} adaptive approach to this problem is still lacking.

To this end, we propose an \textit{online} adaptive filtering algorithm for streaming graph data. Similarly to~\cite{Mei2017}, we design the algorithm in a system identification setting, whereby the task boils down to recovering the structure of the underlying GSO. The proposed approach employs modified stochastic gradient descent methods~\cite{Yim2017,Yim2019}, in addition to graph-specific structures such as sparsity or commutativity, which are enforced naturally by graph topology. As the existing work has already shown the potentials and limits of the graph topology identification problem, this paper aims to further explore the possibility of applying the techniques of adaptive signal processing to random graph processes. The paper is organized as follows. Section II reviews the necessary background. Then, the main problem is outlined in Section III, with the proposed algorithm introduced in Section IV. In Section V, an empirical simulation is performed to validate the capability of the proposed adaptive graph filter. Finally, conclusions are provided in Section VI.
\section{Basics of Graph Signal Processing}
This section reviews the structure of graph signals, system on graphs based on random graph processes, and some relevant notions of weakly stationary graph processes.  

\subsection{Graph Signals}
Consider a weighted random graph, $\mathcal{G}=(\mathcal{V},\mathcal{E},\mathbf{W})$, where the vertex set $\mathcal{V}=\{v_1,v_2,...,v_N\}$, $\mathcal{E}$ is the edge set, and the matrix $\mathbf{W}\in\mathbb{R}^{N\times N}$ is the associated shift operator whose entries $w_{ij}\neq 0$ only if $(i,j)\in\mathcal{E}$. The matrix $\mathbf{W}$ captures the local, usually sparse, patterns of $\mathcal{G}$, the examples of which include (weighted) adjacency matrix, graph Laplacian, and their respective generalized forms~\cite{Stankovic2019,Marques2017}. 

A graph signal is then a function which maps the vertex set, $\mathcal{V}$, onto the set of real or complex numbers, e.g. $f:\mathcal{V}\rightarrow \mathbb{R}$, and is conveniently represented by a vector $\mathbf{x}=[x_1,...,x_N]^T\in\mathbb{R}^N$ where $x_n$ denotes the signal value at vertex $n$. At a particular time instance, the interaction of all elements of a graph signal are modelled according to the graph shift operator (GSO), $\mathbf{W}$, which represents a linear transformation which describes how the graph signals localize across the network.

Similar to the standard shift in time, we can introduce a graph filter which shifts in vertices, $H_L:\mathbb{R}^N\rightarrow\mathbb{R}^N$, defined as a polynomial of graph shift operators in the form
\vspace{-2mm}
\begin{equation}
\label{eq:1}
H_L(\mathbf{W}, \mathbf{h}_k)\triangleq\sum\limits_{l=0}^L{h_{kl} \mathbf{W}^l}
\end{equation}
where $\mathbf{h}_k=[h_{k0},h_{k1},...,h_{kL}]^T$ is a vector of coefficients; the definition of $\mathbf{h}_k$ is given in this way to ease the problem formulation later in the paper. It is noteworthy that the filter $H_L(\mathbf{W}, \mathbf{h}_k)$ is commutative with respect to the shift operator $\mathbf{W}$, that is
\vspace{-1mm}
\begin{equation}
\label{eq:2}
H_K(\mathbf{W}, \mathbf{a})H_L(\mathbf{W}, \mathbf{b})=H_L(\mathbf{W}, \mathbf{b})H_K(\mathbf{W}, \mathbf{a}).
\end{equation}
This property, called the \textit{shift invariance}, will be necessary in the estimation of $\mathbf{W}$ as it implies that the structures of graph processes are not entirely arbitrary.

If $\mathbf{W}$ is an adjacency matrix, then its entries $w_{ij}\geq0$ for $i\neq j$ and $w_{ij}=0$ for $i=j$. When used as a GSO, a graph Laplacian $\mathbf{L}$ (of $\mathbf{W}$) will have a zero row sum with entries $l_{ij}=-w_{ij}$ for $i\neq j$ and $l_{ij}=\sum_iw_{ij}$ for $i=j$. Other alternative GSOs have also been recently proposed~\cite{Bruno2019}, the most suitable choice of which will depend on the application at hand. For example, electric circuits are mainly modelled using adjacency matrices while diffusion-on-graph problems naturally employ the Laplacian. Here, to maintain the generality of this study, the only two assumptions made on $\mathbf{W}$ are the shift invariance and sparsity, common features shared by most GSOs in practice~\cite{Stankovic2019,Marques2017}.

\subsection{Vertex-Time ARMA Processes}
It is important to note that the shift across vertices does not account for the shift in time which reflects the dynamics of real-world signals. Consider a general time-varying $N$-dimensional signal, $\mathbf{x}_t$, generated from another time-varying $N$-dimensional signal, $\mathbf{w}_t$, through a multivariate autoregressive moving average (ARMA) graph process, to give
\begin{equation}
\label{eq:3}
\mathbf{x}_t=\sum\limits_{p=1}^P\boldsymbol{\Psi}_p\mathbf{x}_{t-p}+\sum\limits_{q=0}^Q\boldsymbol{\Phi}_q\mathbf{w}_{t-q}
\end{equation}
where $\boldsymbol{\Psi}_p$ and $\boldsymbol{\Phi}_q$ are coefficient matrices of $\mathbf{x}_t$, so that these matrices are not fixed. For a graph signal, the coefficients (matrix elements) explain how each dimension interacts with all others, and will naturally assume a form of graph shift operators. An intuitive approach would be for the coefficients to assume a form of a graph filter, although there are other interesting basis functions to consider as an alternative, e.g. radial basis functions. Much existing literature~\cite{Lorenzo2016,Lorenzo2018,Qiu2017,Romero2017,Isufi2019,Mei2017} employs polynomial graph filter, also adopted here, whereby $\boldsymbol{\Psi}_p$ and $\boldsymbol{\Phi}_q$ in~\eqref{eq:3} can be expressed as
\vspace{-1mm}
\begin{equation}
\label{eq:4}
\boldsymbol{\Psi}_p\triangleq H_{L_p}(\mathbf{W}, \mathbf{h}_p)
\end{equation}
\begin{equation}
\label{eq:5}
\boldsymbol{\Phi}_q\triangleq H_{K_q}(\mathbf{W}, \mathbf{h}_q)
\end{equation}
where $L_p$ and $K_q$ are integers denoting the maximal shifts of the specific graph filters. With~\cref{eq:4,eq:5}, $\mathbf{x}_t$ and $\mathbf{w}_t$ become graph signals, of which the elements relate to their respective vertex. The values of $L_p$ and $K_q$ are arbitrary and have to be determined for every problem at hand~\cite{Isufi2019}. In this work, we narrow down the scope of the problem by restricting the random graph process to be purely autoregressive and causal~\cite{Mei2017}, thus reducing~\eqref{eq:3} to
\vspace{-1mm}
\begin{equation}
\label{eq:6}
\mathbf{x}_t=\sum\limits_{p=1}^P\boldsymbol{\Psi}_p\mathbf{x}_{t-p}+\mathbf{w}_t
\end{equation}
where $\mathbf{w}_t\sim\mathcal{N}(\mathbf{0},\mathbf{I})$ and \begin{equation}
\label{eq:7}
\boldsymbol{\Psi}_p\triangleq H_p(\mathbf{W}, \mathbf{h}_p).
\end{equation}
\begin{myrmk}
\textnormal{The causality assumption in~\eqref{eq:6} implies that $L_p=p$ and this interpretation is interesting in that the the maximum vertex shifts at a particular time lag cannot exceed the time lag itself. This signifies that \textit{a shift in vertices occurs in tandem with a shift in time i.e. no more than one shift operator is allowed per time instance}. This assumption is rather reasonable as we are analyzing discrete-time models where the sampling policy can be adjusted accordingly.}
\end{myrmk}
\noindent In practice, there may exist a more complicated system where shifts in vertices happen asynchronously with time shift, however small the sampling rate; this is beyond the scope of our work as we believe this scenario is rare. We therefore focus on the vertex-time AR model given in~\eqref{eq:6} and~\eqref{eq:7}.

\subsection{Weak Stationarity}
We shall now briefly explain how the shift invariance of the graph filter can be interpreted as a form of `stationarity'. Analogous to the autocorrelation in time series, the autocorrelation of a graph signal should depend on the `distance' of vertex shifts regardless of the position in vertex domain where the signal initially resides, i.e. however many times the signal has been shifted. Therefore, as long as the total number of shifts is the same, then so should be the correlation. This property is called `weak stationarity' as in Defnition 1 in~\cite{Marques2017}. While the concept of stationarity in graphs is still an open research topic, we employ the notion in~\cite{Marques2017} as it suits our problem setting since the AR model in~\eqref{eq:6} is inherently weakly stationary, as it is made up of a sum of shift-invariant graph filters.

\section{Regularized Least Squares Estimation}
The problem in~\cref{eq:6,eq:7} pertains to the class of multivariate linear regression problems, for which the optimal linear estimator is the MSE estimator~\cite{Papoulis2002}. Here, we adopt the least squares method - a deterministic counterpart of the MSE estimator~\cite{Diniz2012,Yim2017,Yim2019}. The least squares problem of~\cref{eq:6,eq:7} is then given by
\begin{equation}
\label{eq:8}
\underset{\mathbf{W},\mathbf{h}}{\text{min}}\;\;\frac{1}{2}\sum\limits_{\tau=1}^t {\lambda}^{t-\tau}{\left\|\mathbf{x}_{\tau}-\sum\limits_{p=1}^PH_p(\mathbf{W}, \mathbf{h}_p)\mathbf{x}_{{\tau}-p}\right\|}_2^2
\end{equation}
where $\mathbf{h} = [\mathbf{h}_1^T,...,\mathbf{h}_P^T]^T\in\mathbb{C}^M$ with $M={P(P+3)/2}$ and $P$ is the order of this AR random graph process, $\mathbf{x}_i=\mathbf{0}$ for $i\leq 0$ and $\lambda\in(0,1]$. Observe that~\eqref{eq:8} represents a non-convex polynomial problem with many minima, for which many solutions have been proposed, none of which guarantees a global optimum, even under some quite restrictive assumptions~\cite{Mei2017}. A more plausible metric would be therefore to identify whether an edge between any pair of vertices exist with the least chance of misses and false alarms, and the order $P$ should be as small as possible. The above setting implies that $\mathbf{W}$ and $\mathbf{h}$ should be sparse, so that rather than solving the polynomial problem, we can cast~\eqref{eq:8} into alternating steps of regularized least squares sub-problems, outlined in the following sections.

\subsection{Solving for $\boldsymbol{\Psi}_p=H_p(\mathbf{W}, \mathbf{h}_p)$}
The minimization problem in~\cref{eq:8} is now solved with respect to $\boldsymbol{\Psi}_p=H_p(\mathbf{W}, \mathbf{h}_p)$, instead of $\mathbf{W}$ and $\mathbf{h}$; this makes the problem quadratic in $\boldsymbol{\Psi}_p$ and hence standard stochastic gradient descent is applicable. Denote by $\hat{\boldsymbol{\Psi}}_p$ an estimate of $\boldsymbol{\Psi}_p$. With the assumption of sparsity, we now arrive at the optimization problem,
\begin{equation}
\label{eq:9}
\begin{split}
\underset{\boldsymbol{\Psi}}{\text{min}}\;\;\frac{1}{2}&\sum\limits_{\tau=1}^t{\lambda}^{t-\tau}{\left\|\mathbf{x}_{\tau}-\sum\limits_{p=1}^P\boldsymbol{\Psi}_p\mathbf{x}_{{\tau}-p}\right\|}_2^2 \\
&+\sum\limits_{p=1}^P\mu_p{\|\text{vec}(\boldsymbol{\Psi}_p)\|}_1
\end{split}
\end{equation}
where $\boldsymbol{\Psi}=[\boldsymbol{\Psi}_1,...,\boldsymbol{\Psi}_P]\in\mathbb{R}^{N \times NP}$, vec$(\cdot)$ is a vectorization operator and ${\|\cdot\|}_1$ is an $\ell_1$ norm, while $\mu_p$ is a constant which adjusts the degree of sparsity of the corresponding $\boldsymbol{\Psi}_p$. From~\eqref{eq:7}, it is obvious that $\boldsymbol{\Psi}_p$ grows less sparse with an increase in $p$, and thus $\mu_p$ should be set in a decreasing fashion.

The above equation does not account for the shift invariance property of $\boldsymbol{\Psi}_p$. However, from~\eqref{eq:2}, we can add another regularizing term to enforce this constraint, using the following commutator~\cite{Mei2017}
\vspace{-1mm}
\begin{equation}
\label{eq:10}
\left[\boldsymbol{\Psi}_i,\boldsymbol{\Psi}_j\right]\triangleq\boldsymbol{\Psi}_i\boldsymbol{\Psi}_j-\boldsymbol{\Psi}_j,\boldsymbol{\Psi}_i.
\end{equation}
Inserting~\eqref{eq:10} into~\eqref{eq:9} yields
\begin{equation}
\label{eq:11}
\begin{split}
\underset{\boldsymbol{\Psi}}{\text{min}}\;\;&\frac{1}{2}\sum\limits_{\tau=1}^t{\lambda}^{t-\tau}{\left\|\mathbf{x}_{\tau}-\sum\limits_{p=1}^P\boldsymbol{\Psi}_p\mathbf{x}_{{\tau}-p}\right\|}_2^2 \\
&+\sum\limits_{p=1}^P\mu_p{\|\text{vec}(\boldsymbol{\Psi}_p)\|}_1+\gamma\sum\limits_{i \neq j}{\|\left[\boldsymbol{\Psi}_i,\boldsymbol{\Psi}_j\right]\|}_F^2.
\end{split}
\end{equation}
The final term makes the above problem of a quartic programming type, rendering the convergence analysis more difficult. This becomes evident in the simulations where the addition of this regularizer did not improve the algorithm performance significantly, insteadeven slightly deteriorate it in some cases.

\subsection{Estimating $\mathbf{W}$ from $\hat{\boldsymbol{\Psi}}_1$}
From~\eqref{eq:1} and~\eqref{eq:7}, observe that $\boldsymbol{\Psi}_1$ is a linear function of $\mathbf{W}$ and thus its estimate, $\hat{\boldsymbol{\Psi}}_1$, could represent a good estimate of $\mathbf{W}$, that is, $\hat{\mathbf{W}}$. To find a true $\mathbf{W}$ after obtaining $\hat{\boldsymbol{\Psi}}$ from~\eqref{eq:9}, another regularized least squares sub-problem is needed, and given by
\begin{equation}
\label{eq:12}
\begin{split}
\underset{\mathbf{W}}{\text{min}}\;\;&\frac{1}{2}{\left\|\boldsymbol{\Psi}_1-\mathbf{W}\right\|}_2^2+\mu_1{\|\text{vec}(\mathbf{W})\|}_1 \\
&+\gamma\sum\limits_{p=2}^P{\|\left[\mathbf{W},\boldsymbol{\Psi}_p\right]\|}_F^2.
\end{split}
\end{equation}
The rightmost term ensures that $\hat{\mathbf{W}}$ is as commutative as possible with all $\hat{\mathbf{\Psi}}_p$ to ensure the shift invariance property. Note that when~\eqref{eq:11} is employed to calculate $\hat{\boldsymbol{\Psi}}_1$, the shift invariance property has already been enforced so that this optimization sub-problem might be bypassed by setting $\hat{\mathbf{W}}=\hat{\boldsymbol{\Psi}}_1$. The implementation strategy is further elucidated in the simulation section.

\subsection{Estimating $\mathbf{h}$}
After obtaining $\hat{\mathbf{W}}$, the original problem~\eqref{eq:8} turns into a quadratic programming one with respect to $\mathbf{h}$. Also, by assuming that $\mathbf{h}$ is sparse, we can rearrange~\eqref{eq:7} and~\eqref{eq:8} into
\begin{equation}
\label{eq:13}
\underset{\mathbf{h}}{\text{min}}\;\;\frac{1}{2}\sum\limits_{\tau=1}^t{\lambda}^{t-\tau}{\left\|\mathbf{x}_{\tau}-\mathbf{Y}_{\tau}\mathbf{h}\right\|}_2^2+\zeta{\|\mathbf{h}\|}_1
\end{equation}
with
\vspace{-1mm}
\begin{equation}
\label{eq:14}
\mathbf{Y}_t=\left[\mathbf{x}_{t-1},\hat{\mathbf{W}}\mathbf{x}_{t-1},...,\mathbf{x}_{t-P},...,\hat{\mathbf{W}}^P\mathbf{x}_{t-P}\right]
\end{equation}
Note that $\mathbf{Y}_t\in\mathbb{R}^{N \times M}$ contains all possible combination of past $P$ vertex-time instances of the graph signal, $\mathbf{x}_t$. While $M=P(P+3)/2$ appears rather large, in practice, the actual order is quite low, with even $M<N$ a likely case. In addition, this step is optional as our main goal is to recover $\mathbf{W}$.

\section{Adaptive Graph Signal Processing}
We now proceed to build upon the alternating optimization problem detailed in the previous section, to introduce a class of adaptive algorithms based on the optimization criterion in~\eqref{eq:8}, with a sparse solution, called \textit{sparsity-aware adaptive algorithm}. Although many methods, such as $\ell_1$-regularized least mean square~\cite{Chen2009} or oracle algorithm~\cite{Lamare2014}, lead to convergence with competitively small MSE, these solutions are rarely sparse, if not at all~\cite{Taheri2011}, as they do not explicitly zero out the elements of the GSO matrix like their offline counterparts such as basis pursuit. On the other hand, methods like ADMM or ALM have proved valuable in solving offline $\ell_1$-regularized problems, and in our endeavor, as the topic is still in its infancy, we adopt the standard stochastic gradient descent but rewrite our main cost function to naturally enforce the solution to `project' on acceptable values~\cite{Schmidt2007} (the prospect of online ADMM and ALM is promising if the underlying algorithm - as in this paper - is designed to work well). To this end, we re-formulate~\eqref{eq:9} and~\eqref{eq:11} to~\eqref{eq:13} by splitting the desired variables ($\hat{\boldsymbol{\Psi}}$, $\hat{\mathbf{W}}$ and $\hat{\mathbf{h}}$) into their positive and negative parts, that is
\begin{equation}
\label{eq:15}
\hat{\boldsymbol{\Psi}}\triangleq\hat{\boldsymbol{\Psi}}_+-\hat{\boldsymbol{\Psi}}_-,\;\;\hat{\mathbf{W}}\triangleq\hat{\mathbf{W}}_+-\hat{\mathbf{W}}_-,\;\;\hat{\mathbf{h}}\triangleq\hat{\mathbf{h}}_+-\hat{\mathbf{h}}_-
\end{equation}
where $(\cdot)_+\geq\mathbf{0}$ and $(\cdot)_-\geq\mathbf{0}$ contain respectively only the positive and negative parts of $(\cdot)$. Note that if $\mathbf{W}$ is an adjacency matrix, then $\hat{\boldsymbol{\Psi}}_-=\hat{\mathbf{W}}_-=\mathbf{0}$ which makes the problem easier. For the Laplacian, this is much more difficult because while clearly it can be split into the positive on- and negative off-diagonals, the real bottleneck is the zero row sum, an equality constraint which is awkward to solve iteratively as it could involve Lagrangian methods. Since the stucture of GSO vary with applications, we here study the general unconstrained $\mathbf{W}$ for generality and analytic insights.

\subsection{Form of the Algorithm}
Based on~\eqref{eq:15}, the ${\|\cdot\|}_1$ operator can be expressed through a product-weighted sum i.e.
\vspace{-2mm}
\begin{equation}
\label{eq:16}
\begin{split}
&\hat{\boldsymbol{\Psi}}=\text{Tr}\left(\mathbf{1}_{N \times N}\hat{\boldsymbol{\Psi}}_+\right)+\text{Tr}\left(\mathbf{1}_{N \times N}\hat{\boldsymbol{\Psi}}_-\right), \\
&\hat{\mathbf{W}}=\text{Tr}\left(\mathbf{1}_{N \times N}\hat{\mathbf{W}}_+\right)+\text{Tr}\left(\mathbf{1}_{N \times N}\hat{\mathbf{W}}_-\right), \\
&\hat{\mathbf{h}}=\mathbf{1}_N^T\hat{\mathbf{h}}_+-\mathbf{1}_N^T\hat{\mathbf{h}}_-
\end{split}
\end{equation}
where $\text{Tr}(\cdot)$ is a trace operator, and $\mathbf{1}_{N \times N}\in\mathbb{R}^{N \times N}$ and $\mathbf{1}_N\in\mathbb{R}^N$ both have all their elements equal to $1$. These formulae enable us to separate the derivatives with respect to the positive and negative parts, where gradient projection can be used to force invalid values to zero, leading to sparsity as a by-product. We now proceed to minimize~\eqref{eq:11} by calculating the gradient with respect to ${(\hat{\boldsymbol{\Psi}}_p)}_+$, denoted by $\nabla^{(t)}_{{(\hat{\boldsymbol{\Psi}}_p)}_+}$, and given by
\vspace{-2mm}
\begin{equation}
\label{eq:17}
\begin{split}
\nabla^{(t)}_{{(\hat{\boldsymbol{\Psi}}_p)}_+}&=\sum\limits_{\tau=1}^t{\lambda}^{t-\tau}\left(\sum\limits_{k=1}^P\hat{\boldsymbol{\Psi}}_{k,t-1}\mathbf{x}_{{\tau}-k}\mathbf{x}^T_{{\tau}-p}-\mathbf{x}_{\tau}\mathbf{x}^T_{{\tau}-p}\right) \\
&+\mu_{p,t}\mathbf{1}_{N \times N}+\gamma\mathbf{Q}_{p,t}
\end{split}
\end{equation}
where
\vspace{-2mm}
\begin{equation}
\label{eq:18}
\mathbf{Q}_{p,t+1}=\sum\limits_{k=2}^P\left(\left[\hat{\boldsymbol{\Psi}}_{p,t},\hat{\boldsymbol{\Psi}}_{k,t}\right]\hat{\boldsymbol{\Psi}}_{k,t}^T-\hat{\boldsymbol{\Psi}}_{k,t}^T\left[\hat{\boldsymbol{\Psi}}_{p,t},\hat{\boldsymbol{\Psi}}_{k,t}\right]\right).
\end{equation}
Note that $\hat{\boldsymbol{\Psi}}_{p,t}$ denotes $\hat{\boldsymbol{\Psi}}_p$ at the time instant $t$ in the algorithm. Now, let $\hat{\boldsymbol{\Psi}}_t$, $\mathbf{M}_t$, $\mathbf{Q}_t\in\mathbb{R}^{N \times NP}$ be respectively defined as
\begin{equation}
\label{eq:19}
\hat{\boldsymbol{\Psi}}_t\triangleq\left[\hat{\boldsymbol{\Psi}}_{1,t},\hat{\boldsymbol{\Psi}}_{2,t},...,\hat{\boldsymbol{\Psi}}_{P,t}\right]:={\hat{\boldsymbol{\Psi}}}_{+_t}-{\hat{\boldsymbol{\Psi}}}_{-_t},
\end{equation}
\begin{equation}
\label{eq:20}
\mathbf{M}_t\triangleq\left[\mu_{1,t}\mathbf{1}_{N \times N},\mu_{2,t}\mathbf{1}_{N \times N},...,\mu_{P,t}\mathbf{1}_{N \times N}\right],
\end{equation}
\begin{equation}
\label{eq:21}
\mathbf{Q}_t\triangleq\left[\mathbf{Q}_{1,t},\mathbf{Q}_{2,t},...,\mathbf{Q}_{P,t}\right],
\end{equation}
and with the following variables,
\begin{equation}
\label{eq:22}
\mathbf{G}_t\triangleq\hat{\boldsymbol{\Psi}}_{t-1}\mathbf{R}_t-(\mathbf{P}_t-\gamma\mathbf{Q}_t),
\end{equation}
\begin{equation}
\label{eq:23}
\mathbf{R}_t\triangleq\sum\limits_{\tau=1}^t{\lambda}^{t-\tau}\mathbf{x}_{P,{\tau}}\mathbf{x}_{P,{\tau}}^T=\lambda\mathbf{R}_{t-1}+\mathbf{x}_{P,t}\mathbf{x}_{P,t}^T,
\end{equation}
\vspace{-2 mm}
\begin{equation}
\label{eq:24}
\mathbf{P}_t\triangleq\sum\limits_{\tau=1}^t{\lambda}^{t-\tau}\mathbf{x}_{\tau}\mathbf{x}_{P,{\tau}}^T=\lambda\mathbf{P}_{t-1}+\mathbf{x}_t\mathbf{x}_{P,t}^T,
\end{equation}
where $\mathbf{x}_{P,t}\in\mathbb{R}^{NP}$ is given by
\vspace{-2 mm}
\begin{equation}
\label{eq:25}
\mathbf{x}_{P,t}\triangleq{\left[\mathbf{x}^T_{t-1},\mathbf{x}^T_{t-2},...,\mathbf{x}^T_{t-P}\right]}^T.
\end{equation}
We can now express the update for ${\hat{\boldsymbol{\Psi}}}_{+_t}$ as a gradient projection, that is
\begin{equation}
\label{eq:26}
{\hat{\boldsymbol{\Psi}}}_{+_t}={\left({\hat{\boldsymbol{\Psi}}}_{+_{t-1}}-(\mathbf{M}_t+\mathbf{G}_t)(\mathbf{A}_t\otimes\mathbf{I}_{N \times N})\right)}_+
\end{equation}
where $\mathbf{A}_t=\text{diag}(\alpha_1,\alpha_2,...,\alpha_P)\in\mathbb{R}^{P \times P}$ is a diagonal matrix of stepsizes. Similarly, we can obtain the update equation for ${\hat{\boldsymbol{\Psi}}}_{-_t}$ as
\begin{equation}
\label{eq:27}
{\hat{\boldsymbol{\Psi}}}_{-_t}={\left({\hat{\boldsymbol{\Psi}}}_{-_{t-1}}-(\mathbf{M}_t-\mathbf{G}_t)(\mathbf{A}_t\otimes\mathbf{I}_{N \times N})\right)}_+.
\end{equation}
The next step involves finding the shift operator $\hat{\mathbf{W}}$. For example, we can easily let $\hat{\mathbf{W}}_t=\hat{\boldsymbol{\Psi}}_{1,t}$, with the derivation so far based on~\eqref{eq:11} where the commutative property of $\hat{\boldsymbol{\Psi}}$ is already taken into account. Another approach may employ a simplified version of~\eqref{eq:9} with $\mathbf{Q}_t=\mathbf{0}$ for all $t$. Since the commutativity is not enforced in~\cref{eq:9}, but is needed when estimating $\hat{\mathbf{W}}_t$, we repeat the same procedure as in~\eqref{eq:12}, resulting in the following,
\begin{equation}
\label{eq:28}
\hat{\mathbf{W}}_t={\hat{\mathbf{W}}}_{+_t}-{\hat{\mathbf{W}}}_{-_t},
\end{equation}
\begin{equation}
\label{eq:29}
{\hat{\mathbf{W}}}_{+_t}={\left({\hat{\mathbf{W}}}_{+_{t-1}}-\beta_t(\mu_{1,t}\mathbf{1}_{N \times N}+\mathbf{V}_t)\right)}_+
\end{equation}
\begin{equation}
\label{eq:30}
{\hat{\mathbf{W}}}_{-_t}={\left({\hat{\mathbf{W}}}_{-_{t-1}}-\beta_t(\mu_{1,t}\mathbf{1}_{N \times N}-\mathbf{V}_t)\right)}_+
\end{equation}
with
\begin{equation}
\label{eq:31}
\mathbf{V}_t=\hat{\mathbf{W}}_{t-1}-(\hat{\boldsymbol{\Psi}}_{1,t}-\gamma\mathbf{S}_t)
\end{equation}
and
\begin{equation}
\label{eq:32}
\mathbf{S}_t=\sum\limits_{k=2}^P\left(\left[\hat{\mathbf{W}}_{t-1},\hat{\boldsymbol{\Psi}}_{k,t}\right]\hat{\boldsymbol{\Psi}}_{k,t}^T-\hat{\boldsymbol{\Psi}}_{k,t}^T\left[\hat{\mathbf{W}}_{t-1},\hat{\boldsymbol{\Psi}}_{k,t}\right]\right).
\end{equation}
Where $\hat{\mathbf{W}}$ is an adjacency matrix, ${\hat{\boldsymbol{\Psi}}}_{-_t}$ and ${\hat{\mathbf{W}}}_{-_t}$ are both set to zero for all $t$.

Since the objective functions in~\eqref{eq:9} and~\eqref{eq:11} are not pure MSE with regularizing terms, this makes them multi-convex and the solution will thus be biased~\cite{Mei2017} and not optimal in terms of MSE. The whole procedure to this point has been to identify the \textit{causative} elements of an GSO without necessarily their correct values. To this end, we employ an approach known as \textit{debiasing}, where in order to eliminate the regularization biases, we fix the zero entries of the obtained GSO $\hat{\mathbf{W}}_t$, and only optimize the non-zero entries via a least squares cost. It comes with a caveat that, by reducing data sample size, the noise could be distorted from normality, thus affecting the minimal MSE criterion from the outset~\cite{Donoho1995} if the original data is rather noisy, or of insufficiently large size.

The final step, the calculation of $\hat{\mathbf{h}}$, is discretionary as our prime purpose is to identify $\hat{\mathbf{W}}$ and as stated above, the components of $\hat{\mathbf{W}}$ may not be accurately computed due to the biased objectives, leading to even erroneous $\hat{\mathbf{h}}$. On the other hand, if desiring to fully identify the temporal structure of the vertex-time AR process, we can solve for $\hat{\mathbf{h}}$ via~\eqref{eq:13} and~\eqref{eq:14}, but $\hat{\mathbf{W}}$ needs to be \textit{debiased} in order for $\hat{\mathbf{h}}$ to be mathematically meaningful. Unlike the two earlier optimiztion sub-problems, $\hat{\mathbf{h}}$ is not strictly conditioned, and its sparsity constraint aims mainly to render the model succinct. Hence, GAR-LMS~\cite{Taheri2011} is employed to arrive at the update equation of $\hat{\mathbf{h}}$, given by
\begin{equation}
\label{eq:33}
\hat{\mathbf{h}}_t=\hat{\mathbf{h}}_{t-1}+\rho_t\left(\mathbf{C}_t\hat{\mathbf{h}}_{t-1}-\mathbf{u}_t+\eta_t\mathbf{b}_t\right)
\end{equation}
where
\begin{equation}
\label{eq:34}
\mathbf{C}_t=\lambda\mathbf{C}_{t-1}+\mathbf{Y}_t^T\mathbf{Y}_t
\end{equation}
\begin{equation}
\label{eq:35}
\mathbf{u}_t=\lambda\mathbf{u}_{t-1}+\mathbf{Y}_t^T\mathbf{x}_t
\end{equation}
\begin{equation}
\label{eq:36}
\mathbf{b}_t:\;b_{i,t}=\frac{\text{sign}(\hat{h}_{i,t-1})}{\epsilon+\hat{h}_{i,t-1}}
\end{equation}
and
\begin{equation}
\label{eq:37}
\mathbf{Y}_t=\left[\mathbf{x}_{t-1},\hat{\mathbf{W}}_t\mathbf{x}_{t-1},...,\mathbf{x}_{t-P},...,\hat{\mathbf{W}}_t^P\mathbf{x}_{t-P}\right],
\end{equation}
with $\epsilon$ a small positive number. This step could be further simplified by only taking the instantaneous samples into~\eqref{eq:33}, that is, $\lambda=0$, to yield
\begin{equation}
\label{eq:38}
\hat{\mathbf{h}}_t=\hat{\mathbf{h}}_{t-1}+\rho_t\left(\mathbf{Y}_t^T\mathbf{e}_t+\eta_t\mathbf{b}_t\right)
\end{equation}
where
\begin{equation}
\label{eq:39}
\mathbf{e}_t=\mathbf{x}_t-\mathbf{Y}_t\hat{\mathbf{h}}_{t-1}
\end{equation}
The so derived algorithms are summarized in Algorithms 1 \& 2.
\begin{algorithm}
\label{algo:1}
\SetKwInOut{Input}{Input}
\SetKwInOut{Output}{Output}
\SetKwRepeat{DoWhile}{do}{while}

\Input{$\mathbf{x}$, $P$}
\Output{$\hat{\boldsymbol{\Psi}}$, $\hat{\mathbf{W}}^{\ast}$}
 Initialize $\hat{\boldsymbol{\Psi}}_0={\hat{\boldsymbol{\Psi}}}_{+_0}={\hat{\boldsymbol{\Psi}}}_{-_0}=\mathbf{P}_0=\mathbf{Q}_1=\mathbf{0}$, $\hat{\mathbf{W}}_0={\hat{\mathbf{W}}}_{+_0}={\hat{\mathbf{W}}}_{-_0}=\mathbf{S}_1=\mathbf{0}$ and $\mathbf{R}_0=\mathbf{0}$\;
 $t=0$\;
 \DoWhile{$t<T^\ast$ \textnormal{(an epoch with steady state reached)}}{
  $t=t+1$\;
  
  \textit{Solving for} $\hat{\boldsymbol{\Psi}}_t$\;
  $\mathbf{x}_{P,t}={\left[\mathbf{x}^T_{t-1},\mathbf{x}^T_{t-2},...,\mathbf{x}^T_{t-P}\right]}^T$\;
  $\mathbf{R}_t=\lambda\mathbf{R}_{t-1}+\mathbf{x}_{P,t}\mathbf{x}_{P,t}^T$\;
  $\mathbf{P}_t=\lambda\mathbf{P}_{t-1}+\mathbf{x}_t\mathbf{x}_{P,t}^T$\;
  $\mathbf{Q}_t=\left[\mathbf{Q}_{1,t},\mathbf{Q}_{2,t},...,\mathbf{Q}_{P,t}\right]$ with $\mathbf{Q}_{p,t}$ according to~\cref{eq:18}\;
  calculate $\boldsymbol{\mu}_t$\;
  $\mathbf{M}_t=\left[\mu_{1,t}\mathbf{1}_{N \times N},\mu_{2,t}\mathbf{1}_{N \times N},...,\mu_{P,t}\mathbf{1}_{N \times N}\right]$\;
  $\mathbf{G}_t=\hat{\boldsymbol{\Psi}}_{t-1}\mathbf{R}_t-(\mathbf{P}_t-\gamma\mathbf{Q}_t)$\;
  calculate $\mathbf{A}_t$\;
  ${\hat{\boldsymbol{\Psi}}}_{+_t}={\left({\hat{\boldsymbol{\Psi}}}_{+_{t-1}}-(\mathbf{M}_t+\mathbf{G}_t)(\mathbf{A}_t\otimes\mathbf{I}_{N \times N})\right)}_+$\;
  ${\hat{\boldsymbol{\Psi}}}_{-_t}={\left({\hat{\boldsymbol{\Psi}}}_{-_{t-1}}-(\mathbf{M}_t-\mathbf{G}_t)(\mathbf{A}_t\otimes\mathbf{I}_{N \times N})\right)}_+$\;
  $\hat{\boldsymbol{\Psi}}_t={\hat{\boldsymbol{\Psi}}}_{+_t}-{\hat{\boldsymbol{\Psi}}}_{-_t}$\;
  
  \textit{Estimating} $\hat{\mathbf{W}}_t$\;
  $\mathbf{S}_t$ according to~\cref{eq:32}\;
  $\mathbf{V}_t=\hat{\mathbf{W}}_{t-1}-(\hat{\boldsymbol{\Psi}}_{1,t}-\gamma\mathbf{S}_t)$\;
  ${\hat{\mathbf{W}}}_{+_t}={\left({\hat{\mathbf{W}}}_{+_{t-1}}-\beta_t(\mu_{1,t}\mathbf{1}_{N \times N}+\mathbf{V}_t)\right)}_+$\;
  ${\hat{\mathbf{W}}}_{-_t}={\left({\hat{\mathbf{W}}}_{-_{t-1}}-\beta_t(\mu_{1,t}\mathbf{1}_{N \times N}-\mathbf{V}_t)\right)}_+$\;
  $\hat{\mathbf{W}}_t={\hat{\mathbf{W}}}_{+_t}-{\hat{\mathbf{W}}}_{-_t}$\;
 }
 $\hat{\mathbf{W}}^{\ast}=\hat{\mathbf{W}}_{T^\ast}$.
\caption{Identifying the topology of $\hat{\mathbf{W}}$ ($\hat{\mathbf{W}}^{\ast}$)}
\end{algorithm}

As mentioned earlier, two paths are possible for Algorithm 1; either to ignore Step $9$, i.e. $\mathbf{Q}_t=\mathbf{0}$, and consider only Step $18$ (we will call this \textbf{Path 1}), or vice versa - to consider Step $9$ and ignore Step $18$ by letting $\hat{\mathbf{W}}_t=\hat{\boldsymbol{\Psi}}_{1,t}$ (\textbf{Path 2}).

\subsection{Fine-tuning Peripheral Parameters}
For desirable accuracy and \textit{fidelity} of the outcome, there are still some minor parameters which need to be fine tuned. These include the regularization constants $\boldsymbol{\mu}_t:={[\mu_{1,t},\mu_{2,t},...,\mu_{P,t}]}^T$, $\eta_t$, $\gamma$ and $\epsilon$; stepsizes $\mathbf{A}_t$, $\mathbf{\beta}_t$ and $\mathbf{\rho}_t$; and the forgetting factor $\lambda$.

For the $\ell_1$-norm related constants, these can be initially expressed via~\cite{Kim2007}
\begin{equation}
\label{eq:40}
\mu_{p,t}=\mu_p{\left\|\mathbf{P}_{p,t}-\gamma\mathbf{Q}_{p,t}\right\|}_\infty
\end{equation}
\begin{equation}
\label{eq:41}
\eta_t=\eta{\left\|\mathbf{Y}_t^T\mathbf{x}_t\right\|}_\infty
\end{equation}
where $\boldsymbol{\mu}:={[\mu_1,...,\mu_P]}^T$ is a constant vector with entry values decreasing with $p$, and $\mathbf{P}_{p,t}\in\mathbb{R}^{N \times N}$ is a $p^{th}$ block of $\mathbf{P}_t:={\left[\mathbf{P}_{1,t},\mathbf{P}_{2,t},...,\mathbf{P}_{P,t}\right]}$. For the stepsizes, Armijo backtracking is employed to yield suitable values of $\mathbf{A}_t$, $\mathbf{\beta}_t$ and $\mathbf{\rho}_t$, while the parameters $\boldsymbol{\mu}$, $\eta$, $\gamma$ and $\lambda$ have to be determined manually. Notice that while some prior knowledge is available for $\boldsymbol{\mu}$ (decreasing-valued entries) and $\lambda$ (closed to unity), $\eta$ and $\gamma$ are rather unconstrained.

\subsection{Discussion on Convergence}
A rigorous convergence analysis of the graph random processes can be found in~\cite{Mei2017}. However, the assumptions for successful convergence are quite restrictive because the graph signal has not only to obey specific sparsity structure (Assumption \textbf{A5} in~\cite{Mei2017}), but also to exhibit a very strong stability condition (Assumptions \textbf{A4} and \textbf{A6} in~\cite{Mei2017}), to which only a few classes of topologies conform, like $K$-regular graphs. While the proof in~\cite{Mei2017} is without doubt rigorous, it is largely theoretical and limited to real-world cases. Attempts to relax the assumptions underpinning the proof have had limited success; this is partly due to that fact that the base problem~\eqref{eq:9} is inherently biased; for example, the $\ell_1$-norm regularizing terms usually exhibit a side effect of underestimating the non-zero elements~\cite{Kim2007}, not to mention a more complex commutator term. Therefore, it may be more favorable to take a \textit{different} convergence measure. In other words, rather than the mean squared error, we could use the percentage of correctly recovered elements of $\mathbf{W}$, regardless of their correct values, a topic of future work.
\begin{algorithm}
\label{algo:2}
\SetKwInOut{Input}{Input}
\SetKwInOut{Output}{Output}
\SetKwRepeat{DoWhile}{do}{while}
\Input{$\mathbf{x}$, $P$, $\delta$}
\Output{$\hat{\mathbf{W}}$, $\hat{\mathbf{h}}$}
 All recursive variables resume from Algorithm 1\;
 $t=T^\ast$\;
 \DoWhile{$t<T$ \textnormal{(an epoch with $\|\mathbf{e}_T\|<\delta$)}}{
  $t=t+1$\;
  
  \textit{Recovering} $\hat{\mathbf{W}}_t$\;
  $\mathbf{R}_t=\lambda\mathbf{R}_{t-1}+\mathbf{x}_{P,t}\mathbf{x}_{P,t}^T$\;
  $\mathbf{P}_t=\lambda\mathbf{P}_{t-1}+\mathbf{x}_t\mathbf{x}_{P,t}^T$\;
  $\mathbf{G}_t={\left(\hat{\boldsymbol{\Psi}}_{t-1}\mathbf{R}_t-\mathbf{P}_t\right)}_{\hat{\mathbf{W}}}$ where $(\cdot)_{\hat{\mathbf{W}}}$ is the projection to non-zero elements of $\hat{\boldsymbol{\Psi}}$ considering $\hat{\mathbf{W}}$\;
  calculate $\mathbf{A}_t$\;
  ${\hat{\boldsymbol{\Psi}}}_t={\hat{\boldsymbol{\Psi}}}_{t-1}-\mathbf{G}_t(\mathbf{A}_t\otimes\mathbf{I}_{N \times N})$\;
  Setting $\hat{\mathbf{W}}_t=\hat{\boldsymbol{\Psi}}_{1,t}$\;
  
  \textit{Estimating} $\hat{\mathbf{h}}$\;
  $\mathbf{Y}_t=\left[\mathbf{x}_{t-1},\hat{\mathbf{W}}_t\mathbf{x}_{t-1},...,\mathbf{x}_{t-P},...,\hat{\mathbf{W}}_t^P\mathbf{x}_{t-P}\right]$\;
  $\mathbf{e}_t=\mathbf{x}_t-\mathbf{Y}_t\hat{\mathbf{h}}_{t-1}$\;
  $\mathbf{b}_t:\;b_{i,t}=\frac{\text{sign}(\hat{h}_{i,t})}{\sigma+\hat{h}_{i,t}}$\;
  $\hat{\mathbf{h}}_t=\hat{\mathbf{h}}_{t-1}+\rho_t\left(\mathbf{Y}_t^T\mathbf{e}_t+\eta_t\mathbf{b}_t\right)$\;
 }
 $\hat{\mathbf{W}}=\hat{\mathbf{W}}_T$, $\hat{\mathbf{h}}=\hat{\mathbf{h}}_T$.
\caption{Determining the \textit{unbiased} $\hat{\mathbf{W}}$ and $\hat{\mathbf{h}}$}
\end{algorithm}
\section{Simulation Results}
Since the suitable choice of the GSOs varies with applications, in our simulation, we tested our algorithm with synthetic graph processes which are consistent with the underlying assumptions of sparsity and shift invariance. Three different topologies of graphs were considered: arbitrarily random graph (R), random graph with power-law degree distribution (PL)~\cite{Barabasi1999}, and Stochastic Block-Model (SBM)~\cite{Karrer2011}. For each topology, the synthetic graph signal $\mathbf{x}_t\in\mathbb{R}^{12\times12}$ was generated by feeding an i.i.d. input signal $\mathbf{w}_t\in\mathbb{R}^{12\times12}$ into the stochastic processes in~\eqref{eq:6} and~\eqref{eq:7}, with the number of vertices $N=12$ and time lag order $P=3$ throughout all simulations.

\subsection{Convergence Performance against NMSE}
In the first experiment, we examined how the overall algorithm performs in terms of the \textit{normalized mean squared error} (NMSE) of $\mathbf{x}$ and $\mathbf{W}$, respectively denoted by
\begin{equation}
\label{eq:42}
\sigma_t\triangleq\frac{{\|\mathbf{e}_t\|}^2_2}{{\|\mathbf{x}_t\|}^2_2},
\end{equation}
\begin{equation}
\label{eq:43}
\zeta_t\triangleq\frac{{\|\mathbf{W}-\hat{\mathbf{W}}_t\|}^2_F}{{\|\mathbf{W}\|}^2_F}
\end{equation}
where ${\|\cdot\|}_F$ indicates the Frobenius norm. The GSO, $\mathbf{W}$, was generated following the R/PL/SBM topologies chosen at random with 20 realizations in total. For an arbitrarily random topology, the weighted edges were drawn from $\mathcal{N}(0,1)$ and then thresholded to between $0.3$ and $0.7$ times the maximum absolute value of the components. Finally, the GSO matrix was normalized by $1.5$ times its largest eigenvalue (to ensure stable processes).

The PL topology started from three random initial nodes connecting one another with probability $0.8$; then new nodes were connected with the probability following the preferential attachment process~\cite{Barabasi1999} which is proportional to the total weight of the existing nodes. If connected, the weighted edges were drawn from $\mathcal{N}(0,1)$ and thresholded to between $0.05$ and $0.95$ times the maximum absolute value of the components, together with normalizing the GSO matrix by $1.5$ times its largest eigenvalue. In the SBM case, the network was clustered into $3$ groups of $3,4$ and $5$ vertices each. The inter-/intra-cluster probability of connection was allocated by $3\times3$ matrix in the form of $0.25\mathbf{I}+\mathcal{U}(0.05,0.2)$. Then, all the assigned edges were weighted by an exponential distribution with the rate $\lambda=2$ and the matrix was finally normalized by $1.5$ times its largest eigenvalue. All these specifications yielded the sparsity in $\mathbf{W}$ of approximately $0.2$.

After $\mathbf{W}$ was created, $\mathbf{x}_t$ was obtained with the coefficients $h_{ij}$ for $1\leq i\leq P$ and $0\leq j \leq i$, generated sparsely from a mixture of distribution $h_{ij}\sim\frac{1}{2^{i+j}}(\mathcal{U}(-1,-0.45)+\mathcal{U}(0.45,1))$. The data was created for over 1100 samples, with first 500 samples left out due to their transient behavior, and the latter 600 samples kept for the simulation. We employed Algorithm 1 (\textbf{Path 1}) for the first 400 samples and Algorithm 2 for the remaining 200 samples, to recover $\mathbf{x}$ and $\mathbf{W}$, with the hyper-parameters $\mu_1$, $\mu_2$, $\mu_3$, $\eta$ chosen from the interval $(0,5]$ with the step $0.1$, $\gamma$ from $(0,2]$ with the step $0.1$ and $\lambda$ from $(0.8,0.99]$ with the step $0.01$. For this specific experiment, the selected hyperparameters would minimize the steady-state $\sigma_t$ i.e. the averaged $\sigma_t$ for $t$ such that $\sigma_t$ is in steady state. This step was repeated 20 times to obtain 20 realizations which were then averaged to display the outcome.

In terms of NMSE, the regularized algorithm (Algorithm 1) failed to minimize the `normed' error of both $\mathbf{x}$ and $\mathbf{W}$, diverging away and levelling at a certain level; a jittery pattern was observed in the NMSE of $\mathbf{W}$. Afterwards, the debiasing process (Algorithm 2) managed to significantly reduce the error to a very low level, as expected from a generic adaptive algorithm. At the first glance, one may question the utility of the first step (Algorithm 1) as the standard stochastic algorithm (Algorithm 2) can accomplish the whole task. The answer is that Algorithm 2 (debiasing) only manipulates the explanatory part of the $\mathbf{W}$, determined by Algorithm 1. Therefore, it would be a disadvantage to neglect the \textit{capability of identifying the correct topology of} $\mathbf{W}$, which is considered in the next experiment.

\subsection{Identifiability of the GSO Topologies}
The same data formulation as in the previous section was used Here, we focus on the likelihood that Algorithm 1 would successfully identify the right topology i.e. the non-zero elements of $\mathbf{W}$, regardless of their correct values. To this end, we compared the rate of \textit{false alarm} (taking zero element as non-zero) and \textit{miss} (failing to identify non-zero element) for every specific topology. Each case was calculated based on the average of $10$ repeated random trials.
\vspace{-0.2cm}
\begin{figure}[H]
    \centering
    \includegraphics[width=\columnwidth]{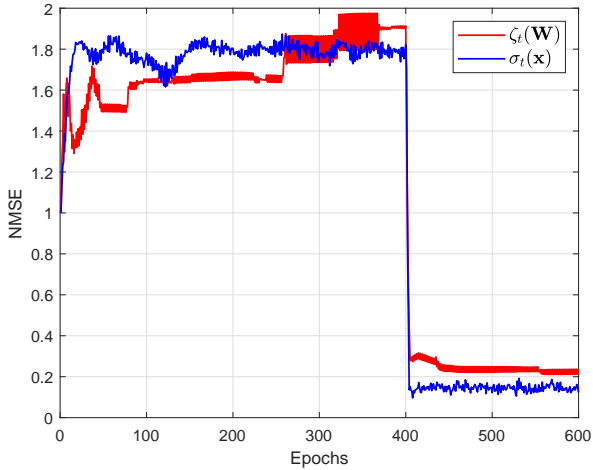}
    \caption{NMSE performance of the proposed algorithm with measures $\zeta_t$ (red) and $\sigma_t$ (blue) which represent respectively the NMSE of $\mathbf{W}$ and $\mathbf{x}$. Algorithm 1 was implemented for the first 400 epochs of data samples and Algorithm 2 for the last 200 epochs.}
    \label{fig:1}
\end{figure}
\textit{As a consistent benchmark of the outcome}, we tuned the hyperparameters such that, in each simulation, the sum of total false alarms ($P_{FA}$) and total misses ($P_M$) was minimal with respect to the hyperparameter grids described in the previous experiment. We tested our data twice with the \textbf{Path 1} and the \textbf{Path 2} of Algorithm 1, to establish if this shifting order of the commutator term affects the learning performance. Table I shows the probabilities of false alarm ($P_{FA}$) and miss ($P_M$) via Path 1 of Algorithm 1 while Table II shows those of Path 2. Observe that Path 1 (using commutator term when estimating
\begin{table}[!htb]
\begin{minipage}{.5\linewidth}
    \centering
    \label{tab:first_table}
    \medskip
\begin{tabular}{ c  c  c }
    \toprule
    \makecell{$\mathbf{W}$}  & $P_{FA}$ & $P_M$ \\ 
    \midrule
    random & $0.1033$ & $0.1967$ \\
    SBM & $0.02$ & $0.353$ \\
    PL & $0.067$ & $0.42$ \\
\bottomrule
\end{tabular}
\caption{Results for Path 1}
\end{minipage}\hfill
\begin{minipage}{.5\linewidth}
    \centering
    \label{tab:second_table}
    \medskip
 \begin{tabular}{ c  c  c }
    \toprule
    \makecell{$\mathbf{W}$}  & $P_{FA}$ & $P_M$ \\ 
    \midrule
    random & $0.22$ & $0.2167$ \\
    SBM & $0.0633$ & $0.4567$ \\
    PL & $0.15$ & $0.6$ \\
\bottomrule
\end{tabular}
\caption{Results for Path 2}
\end{minipage}
\end{table}
$\hat{\mathbf{W}}_t$ rather than at solving for $\hat{\boldsymbol{\Psi}}_t$) yielded more accurate recovery than the Path 2 (using the commutator term together with solving for $\hat{\boldsymbol{\Psi}}_t$ and letting $\hat{\mathbf{W}}_t=\hat{\boldsymbol{\Psi}}_{1,t}$). Regarding topology-wise comparison, the results expectedly show that specific topologies affect algorithm performance. When testing the arbitrarily random topology, we noticed the resulting recovery was \textit{not} consistent, as indicated by the sum of $P_{FA}$ and $P_M$ varying considerably from experiment to experiment. Fig 2 (a) shows one of random-topology trials which are close to the average. When considering $\mathbf{W}$ with a clearly specified topology (SBM and PL), the recovery rate was more consistent with most SBM and PL trials, giving the recovery outcome close to the average. Fig 2 (b) and (c) visualize trial cases for both topologies.
\begin{figure}[!htb]
    \centering
    \begin{subfigure}[ht]{0.32\columnwidth}
        \includegraphics[width=\linewidth]{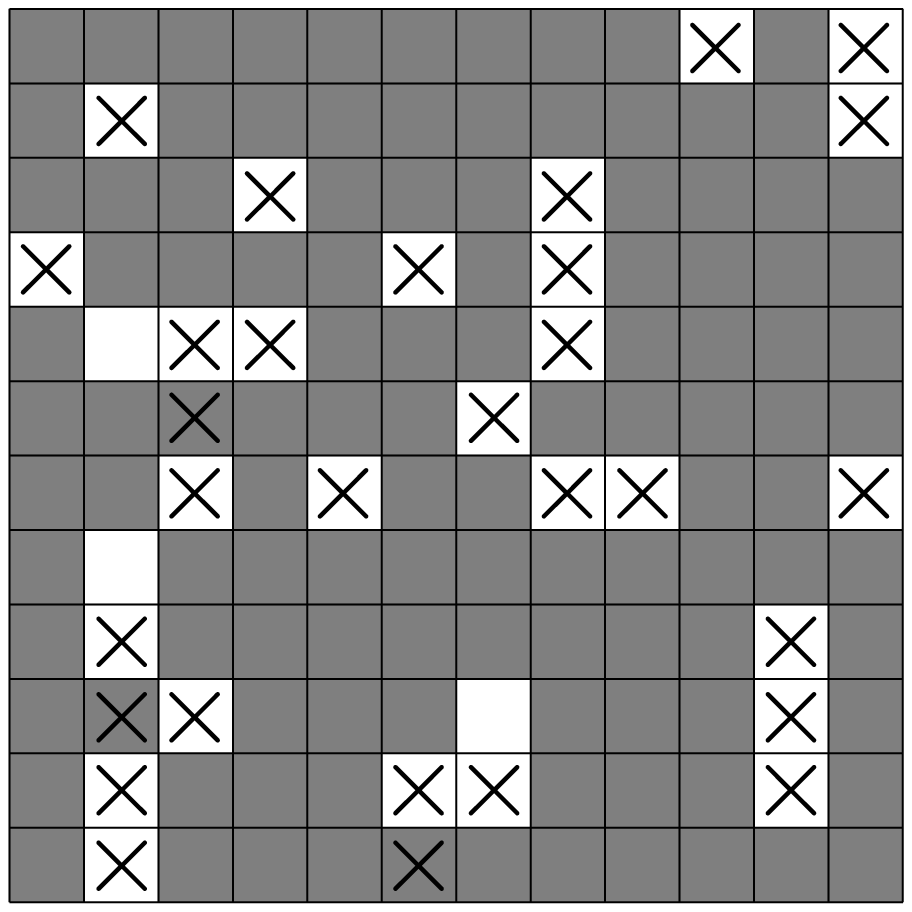}
        \caption{}
        \label{fig:2a}
    \end{subfigure}
    \begin{subfigure}[ht]{0.32\columnwidth}
        \includegraphics[width=\linewidth]{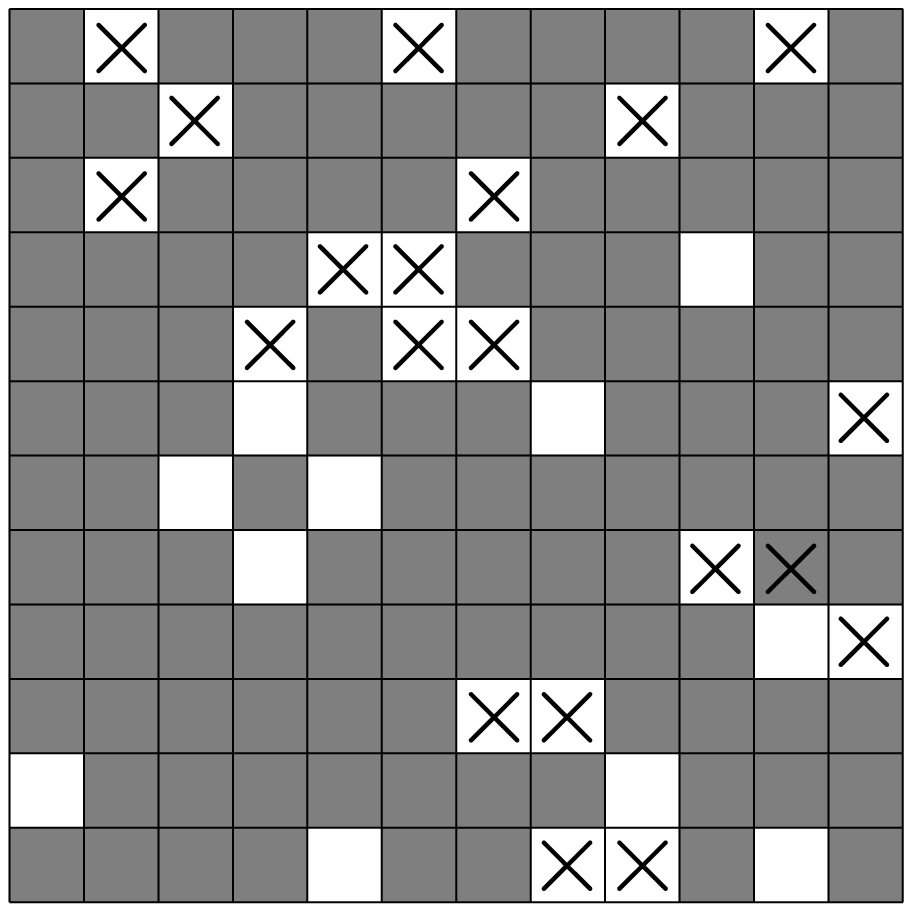}
        \caption{}
        \label{fig:2b}
    \end{subfigure}
    \begin{subfigure}[ht]{0.32\columnwidth}
        \includegraphics[width=\linewidth]{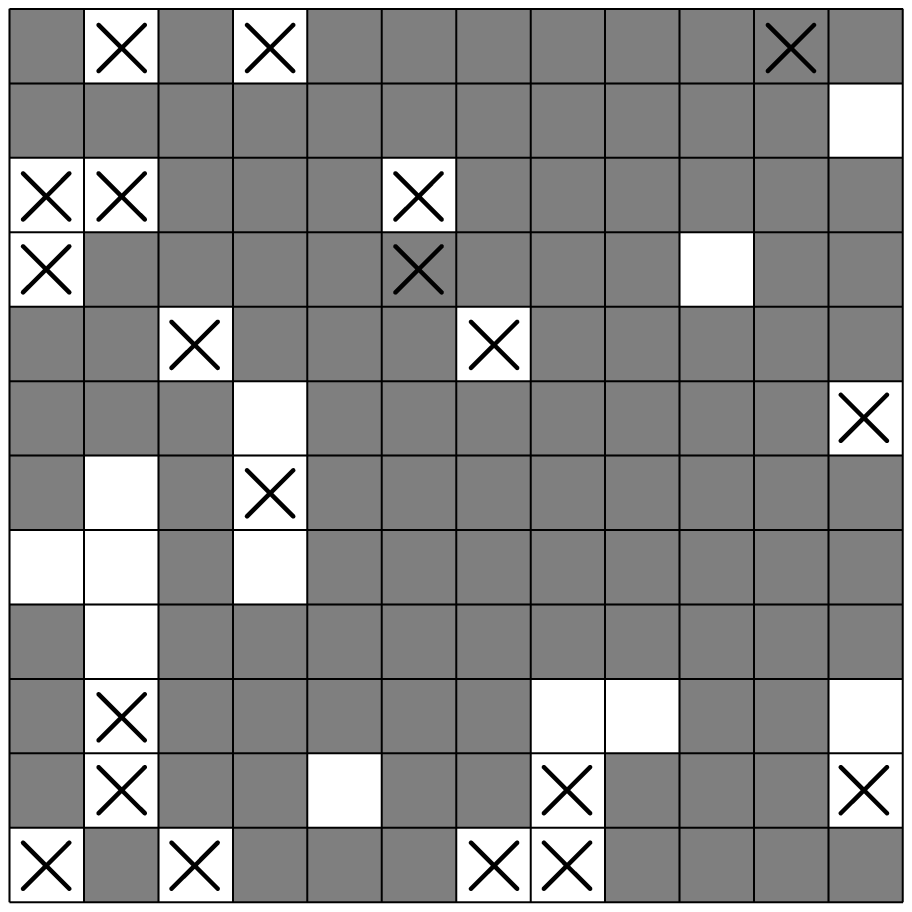}
        \caption{}
        \label{fig:2c}
    \end{subfigure}
    \caption{Examples of visual representation of (a) arbitrarily random, (b) SBM and (c) power-law topologies of the graph shift operator, $\mathbf{W}$, from one simulation trial where blank spaces designate the non-zero entries, grey spaces the zero entries, and crosses the recovered entries.}
    \label{fig:2}
\end{figure}
It should be noted that the recovery results of SBM and PL topologies were not outstanding as some trails of the arbitrarily random topology gave more precise identification; an example is shown in Fig. 2. While the structure of SBM and PL graphs ensured that the algorithm was less susceptible to identifying wrong edges, they disproportionately lacked in the ability to detect all the right ones (their $P_M$'s were quite high compared to the \textit{random} benchmark). When attempting to reduce high $P_M$'s, their $P_{FA}$ outgrew the intended reduction; in other words, after reaching some \textit{point}, the algorithm began to wrongly identify edges at a rapid rate. Visually, we still could not distinguish what characteristics of $\mathbf{W}$ would render the algorithm more effective.

Finally, we would like to mention that these simulations were run on a small-scale problem due to computational limits, and hence the $12\times 12$ size of $\mathbf{W}$ and $10$ trails per topology could give biased and more varying results compared with the experiments involving thousands of nodes and hundreds of trials~\cite{Mei2017}. Nevertheless, the findings in this work indicate that there is much more room to discover in this research topic, since topology constraints play a crucial role in selecting appropriate optimization techniques to devise learning algorithms. This already suggests that other topological statistics like graph diameter, node degrees and many others could help with the design of the optimal algorithms.
\section{Conclusion}
This paper presents a first design of adaptive graph signal processing implemented jointly by formulating a problem and devising a novel online algorithm accordingly. The model is based on vector autoregression (VAR) where the coefficient matrices are constrained by graph topology via a graph shift operator (GSO). The vertex-time relationship has been explained through a graph filter (vertex part) embedded into a VAR time series (time part), where causality has been imposed on the model to determine lag characteristics of the vertex-time models. To alleviate the non-convex nature arising from the polynomial structure of the graph filters, the problem has been divided into three sub-problems which themselves are re-expressed as convex problems. The algorithm has then been derived based on the split gradient projection method~\cite{Schmidt2007} which groups the first two sub-problems into Algorithm 1 and the last into Algorithm 2. The reason is that the method is expected to produce biased results due to heavy-handed regularization of the problem. Therefore, after Algorithm 1, only the non-zero entries of the resulting GSO are computed in Algorithm 2 to \textit{debias} the solution. Finally, the experiments have been carried out to illustrate the potentials and limits of the proposed method. The fine-tuning of hyperparameters poses another challenge as the empirical results from the algorithm is shown to be highly susceptible to how these hyperparameters are collectively set.


%




\ifCLASSOPTIONcaptionsoff
  \newpage
\fi


\begin{thebibliography}{1}

\bibitem{Chambers2001}
D. P. Mandic and J. Chambers. 2001. {\em Recurrent Neural Networks for Prediction: Learning Algorithms, Architectures and Stability}. John Wiley \& Sons, Inc., New York, NY, USA.

\bibitem{Diniz2012}
P. S. R. Diniz, \emph{Adaptive Filtering: Algorithms and Practical Implementation}, Springer Science \& Business Media, 2012.

\bibitem{Tao2013}
J. W. Tao and W. X. Chang, ``A novel combined beamformer based on hypercomplex processes," \emph{IEEE Trans. Aero. Electron. Sys.}, vol. 49, no. 2, pp. 1276-1289, Apr. 2013.

\bibitem{Yim2017}
T. Variddhisa\"i and D. P. Mandic, ``On an RLS-like LMS adaptive filter,'' in {\em Proc. 22nd Int. Conf. Digital Signal Process. (DSP)}, London, 2017, pp. 1-5.

\bibitem{Took2010}
C. Took and D. P. Mandic, ``A quaternion widely linear adaptive filter," \emph{IEEE Trans. Signal Process.}, vol. 58, no. 8, pp. 4427-4431, Aug. 2010.

\bibitem{Yim2019}
T. Variddhisa\"i and D. P. Mandic, ``Online Multilinear Dictionary Learning,'' {\em arXiv}, 1703.02492, cs.LG, 2017

\bibitem{Widrow1959}
B. Widrow, M.E. Hoff, ``Adaptive switching circuits," in \emph{IRE Wescon Conv Rec 4}, pp. 96–104.

\bibitem{Lorenzo2016}
P. Di Lorenzo, S. Barbarossa, P. Banelli and S. Sardellitti, ``Adaptive Least Mean Squares Estimation of Graph Signals,'' \emph{IEEE Trans. Signal Info. Process. Net.}, vol. 2, no. 4, pp. 555-568, Dec. 2016.

\bibitem{Lorenzo2018}
P. Di Lorenzo, P. Banelli, E. Isufi, S. Barbarossa and G. Leus, ``Adaptive Graph Signal Processing: Algorithms and Optimal Sampling Strategies,'' \emph{IEEE Trans. Signal Process.}, vol. 66, no. 13, pp. 3584-3598, Jul. 2018.

\bibitem{Qiu2017}
K. Qiu, X. Mao, X. Shen, X. Wang, T. Li and Y. Gu, ``Time-Varying Graph Signal Reconstruction,'' \emph{IEEE Jour. Select. Topic. Sig. Process.}, vol. 11, no. 6, pp. 870-883, Sep 2017.

\bibitem{Romero2017}
D. Romero, V. N. Ioannidis and G. B. Giannakis, ``Kernel-Based Reconstruction of Space-Time Functions on Dynamic Graphs,'' \emph{IEEE Jour. Select. Topic. Sig. Process.}, vol. 11, no. 6, pp. 856-869, Sep 2017.

\bibitem{Isufi2019}
E. Isufi, A. Loukas, N. Perraudin and G. Leus, ``Forecasting Time Series With VARMA Recursions on Graphs,'' \emph{IEEE Trans. Signal Process.}, vol. 67, no. 18, pp. 4870-4885, Sep 2019.

\bibitem{Mei2017}
J. Mei and J. M. F. Moura, ``Signal Processing on Graphs: Causal Modeling of Unstructured Data,'' \emph{IEEE Trans. Signal Process.}, vol. 65, no. 8, pp. 2077-2092, Apr 2017.

\bibitem{Stankovic2019}
L. Stankovic, D. P. Mandic, M. Dakovic, I. Kisil, E. Sejdic and A. G. Constantinides, ``Understanding the Basis of Graph Signal Processing via an Intuitive Example-Driven Approach [Lecture Notes],'' \emph{IEEE Sig. Process. Mag.}, vol. 36, no. 6, pp. 133-145, Nov 2019.

\bibitem{Marques2017}
A. G. Marques, S. Segarra, G. Leus and A. Ribeiro, ``Stationary Graph Processes and Spectral Estimation,'' \emph{IEEE Trans. Signal Process.}, vol. 65, no. 22, pp. 5911-5926, Nov 2017.

\bibitem{Bruno2019}
B. Scalzo-Dees, L. Stankovic, M. Dakovic, A. G. Constantinides and D. P. Mandic, ``A Class of Doubly Stochastic Shift Operators for Random Graph Signals and their Boundedness,'' {\em arXiv}, 1908.01596, eess.SP, 2019

\bibitem{Papoulis2002}
A. Papoulis, \emph{Probability, Random Variables and Stochastic Processes}. McGraw-Hill Education; 4th edition 2002.

\bibitem{Chen2009}
Y. Chen, Y. Gu and A. O. Hero, ``Sparse LMS for system identification,'' in Proc. \emph{IEEE Int. Conf. Acoust., Speech, Signal Process. (ICASSP)}, Taipei, pp. 3125-3128, 2009.

\bibitem{Lamare2014}
R. C. de Lamare and R. Sampaio-Neto, ``Sparsity-Aware Adaptive Algorithms Based on Alternating Optimization and Shrinkage,'' \emph{IEEE Signal Process. Lett.}, vol. 21, no. 2, pp. 225-229, Feb 2014.

\bibitem{Taheri2011}
O. Taheri and S. A. Vorobyov, ``Sparse channel estimation with lp-norm and reweighted l1-norm penalized least mean squares,'' in Proc. \emph{IEEE Int. Conf. Acoust., Speech, Signal Process. (ICASSP)}, Prague, 2011, pp. 2864-2867.

\bibitem{Schmidt2007}
M. Schmidt, G. Fung, R. Rosales, \emph{Fast Optimization Methods for L1 Regularization: A Comparative Study and Two New Approaches}. Lecture Notes in Computer Science, vol 4701. Springer, Berlin, Heidelberg, 2007.

\bibitem{Donoho1995}
D. L. Donoho, ``De-noising by soft-thresholding,'' \emph{IEEE Trans. Info. Theo.}, vol. 41, no. 3, pp. 613-627, May 1995.

\bibitem{Kim2007}
S. Kim, K. Koh, M. Lustig, S. Boyd, and D. Gorinvesky, ``A Method for Large-Scale $\ell_1$-Regularized Least Squares Problems With Applications in Signal Processing and Statistics,'' Dept. Elect. Eng., Stanford Univ., 2007 [Online].

\bibitem{Barabasi1999}
A. Barabasi and R. Albert, ``Emergence of Scaling in Random Networks,'' \emph{Science}, vol. 286, no. 5439, pp. 509–512, Oct 1999.

\bibitem{Karrer2011}
B. Karrer and M. E. J. Newman, ``Stochastic blockmodels and community structure in networks,'' \emph{Phys. Rev. E, Stat., Nonlin., Soft Matt. Phys.}, vol. 83, no. 1 Pt 2, pp. 016-107, Jan 2011.
\end{thebibliography}
\end{document}